\documentclass[english,aps,reprint,superscriptaddress,prc,nofootinbib]{revtex4-1}
\usepackage[latin9]{inputenc}
\usepackage{color}
\usepackage{textcomp}
\usepackage{amsmath}
\usepackage{graphicx}
\usepackage{subscript}

\makeatletter
 
 \@ifundefined{textcolor}{}
 {%
   \definecolor{BLACK}{gray}{0}
   \definecolor{WHITE}{gray}{1}
   \definecolor{RED}{rgb}{1,0,0}
   \definecolor{GREEN}{rgb}{0,1,0}
   \definecolor{BLUE}{rgb}{0,0,1}
   \definecolor{CYAN}{cmyk}{1,0,0,0}
   \definecolor{MAGENTA}{cmyk}{0,1,0,0}
   \definecolor{YELLOW}{cmyk}{0,0,1,0}
 }

\makeatother

\usepackage{babel}
\begin{document}
\begin{abstract}
The applicability of Coulomb dissociation reactions to determine the cross section for the inverse neutron capture reaction was explored using the reaction $^8$Li($\gamma$,n)$^7$Li. A 69.5 MeV/nucleon $^8$Li beam was incident on a Pb target, and the outgoing neutron and $^7$Li nucleus were measured in coincidence. The deduced (n,$\gamma$) excitation function is consistent with data for the direct capture reaction $^7$Li(n,$\gamma$)$^8$Li and with low-energy effective field theory calculations.
\end{abstract}

\title{Determining the $^7$Li(n,$\gamma$) cross section via Coulomb dissociation of $^8$Li}

\author{R. Izs\'ak}
\affiliation{Department of Atomic Physics, E\"otv\"os Lor\'and University, H-1117 Budapest, Hungary}

\author{\'A. Horv\'ath}
\affiliation{Department of Atomic Physics, E\"otv\"os Lor\'and University, H-1117 Budapest, Hungary}

\author{\'A. Kiss}
\affiliation{Department of Atomic Physics, E\"otv\"os Lor\'and University, H-1117 Budapest, Hungary}

\author{Z. Seres}
\affiliation{Institute for Particle and Nuclear Physics, Wigner Research Centre for Physics, H-1525 Budapest, Hungary}

\author{A. Galonsky}
\affiliation{National Superconducting Cyclotron Laboratory, Michigan State University,
East Lansing, MI 48824, USA}

\author{C.A. Bertulani}
\affiliation{Department of Physics and Astronomy, Texas A\&M University-Commerce,
Commerce, TX 75429, USA}

\author{Zs. F\"ul\"op}
\affiliation{ATOMKI Institute for Nuclear Research, H-4001 Debrecen, Hungary}

\author{T. Baumann}
\affiliation{National Superconducting Cyclotron Laboratory, Michigan State University,
East Lansing, MI 48824, USA}

\author{D. Bazin}
\affiliation{National Superconducting Cyclotron Laboratory, Michigan State University,
East Lansing, MI 48824, USA}

\author{K. Ieki}
\affiliation{Department of Physics, Rikkyo University, 3 Nishi-Ikebukuro, Toshima, Tokyo 171, Japan}

\author{C. Bordeanu}
\altaffiliation{On leave from: Department of Nuclear Physics, ``Horia Hulubei - National Institute for Physics and Nuclear Engineering'', Str. Reactorului 30, Magurele, Jud. Ilfov, 077125, Romania}
\affiliation{Department of Physics, University of Washington, Seattle, WA 98195, USA}

\author{N. Carlin}
\affiliation{Instituto de F\'isica, Universidade de S\~ao Paulo, 05315-970 S\~ao Paulo, Brazil}

\author{M. Csan\'ad}
\affiliation{Department of Atomic Physics, E\"otv\"os Lor\'and University, H-1117 Budapest, Hungary}

\author{F. De\'ak}
\affiliation{Department of Atomic Physics, E\"otv\"os Lor\'and University, H-1117 Budapest, Hungary}

\author{P. DeYoung}
\affiliation{Department of Physics and Engineering, Hope College, Holland, MI 49423, USA}

\author{N. Frank}
\altaffiliation{Permanent address: Department of Physics \& Astronomy, Augustana College, Rock Island, IL 61201, USA}
\affiliation{National Superconducting Cyclotron Laboratory, Michigan State University,
East Lansing, MI 48824, USA}
\affiliation{Department of Physics and Astronomy, Michigan State University, East
Lansing, MI 48824, USA}

\author{T. Fukuchi}
\altaffiliation{Present address: RIKEN Center for Life Science Technologies, Kobe, Hyogo 650-0047, Japan}
\affiliation{Department of Physics, Rikkyo University, 3 Nishi-Ikebukuro, Toshima, Tokyo 171, Japan}

\author{A. Gade}
\affiliation{National Superconducting Cyclotron Laboratory, Michigan State University,
East Lansing, MI 48824, USA}
\affiliation{Department of Physics and Astronomy, Michigan State University, East
Lansing, MI 48824, USA}

\author{D. Galaviz}
\altaffiliation{Present address: Centro de F\'isica Nuclear da Universidade de Lisboa, 1649-003 Lisbon, Portugal}
\affiliation{National Superconducting Cyclotron Laboratory, Michigan State University,
East Lansing, MI 48824, USA}

\author{C. R. Hoffman}
\altaffiliation{Permanent address: Physics Division, Argonne National Laboratory, Argonne, Illinois 60439, USA}
\affiliation{Department of Physics, Florida State University, Tallahassee, FL
32306, USA}

\author{W.A. Peters}
\altaffiliation{Present address: Joint Institute for Nuclear Physics and Applications, Oak Ridge, TN 37831, USA}
\affiliation{National Superconducting Cyclotron Laboratory, Michigan State University,
East Lansing, MI 48824, USA}
\affiliation{Department of Physics and Astronomy, Michigan State University, East
Lansing, MI 48824, USA}

\author{H. Schelin}
\altaffiliation{Permanent address: Pele Pequeno Principe Research Institute, 80250-200, Curitiba-PR, Brazil}
\affiliation{Federal University of Technology - Parana, 80230-901 Curitiba, Paran\'a, Brazil}



\author{M. Thoennessen}
\affiliation{National Superconducting Cyclotron Laboratory, Michigan State University,
East Lansing, MI 48824, USA}
\affiliation{Department of Physics and Astronomy, Michigan State University, East
Lansing, MI 48824, USA}

\author{G.I. Veres}
\affiliation{Department of Atomic Physics, E\"otv\"os Lor\'and University, H-1117 Budapest, Hungary}

\maketitle

\section{Introduction}

It is well established that Coulomb dissociation cross sections can provide
nuclear structure information about neutron-rich nuclei \cite{Nakamura-LectureNotes,Nakamura-FewBody}.
In nuclear astrophysics, neutron capture cross sections ($\sigma_{\textrm{n},\gamma}$) on radioactive
nuclei are important in nucleogenesis, and a cross-section measurement of the inverse
reaction, Coulomb dissociation \cite{Baur1986-188,Baur2001-99}, might
be the only way to obtain the capture cross sections. 
Complementary indirect techniques to determine neutron-capture cross sections, such as so-called ``surrogate'' methods, have been utilized in other systems, but comparison to Coulomb dissociation as well as direct measurements is desirable to better understand the applicability of such approaches  \cite{RevModPhys.84.353}.
In Coulomb dissociation,
the projectile is dissociated into a neutron and a remainder fragment
by a ``target'' photon absorbed from the electric field of a high-\textsf{$Z$} target nucleus. First-order perturbation theory gives the relationship between
the Coulomb dissociation function $d\sigma_{\textrm{CD}}/dE_{\gamma}$
and the photo disintegration cross section $\sigma_{\gamma,\textrm{n}}$,
as \cite{Baur1986-188}
\begin{equation}
\sigma_{\gamma,\textrm{n}}(E_{\gamma},E\lambda)=\frac{E_{\gamma}}{n(E_{\gamma},E\lambda)}\frac{d\sigma_{\textrm{CD}}(E_{\gamma},E\lambda)}{dE_{\gamma}},\label{eq:virtual-photon-method}
\end{equation}
where $n(E_{\gamma},E\lambda)$ is the number of virtual photons
with energy $E_{\gamma}$ and multipolarity $E\lambda$.
The principle of detailed balance \cite{SachsNuclearTheory1953} then
yields $\sigma_{\textrm{n},\gamma}$ from $\sigma_{\gamma,\textrm{n}}$.
It is desirable to test the accuracy of perturbation theory by
comparing $\sigma_{\textrm{n},\gamma}$ values deduced from Coulomb
dissociation with directly-measured values of $\sigma_{\textrm{n},\gamma}$. 

Coulomb dissociation has been extensively applied to extract proton capture cross sections; see Ref.~\cite{Bertulani2010195} for an overview of the various reactions. In contrast, neutron capture cross sections have been deduced from Coulomb dissociation and compared with the direct process only for the system $^{14}$C(n,$\gamma$)$^{15}$C \cite{Horvath2002-926,Nakamura2009-035805}. In this case the dominant process is the capture of $p$-wave neutrons. Nakamura {\it et al.\ }\cite{Nakamura2009-035805} have demonstrated that the cross section derived from Coulomb dissociation of $^{15}$C agrees well with the directly measured capture cross section \cite{PhysRevC.77.015804}.

In the present work we report on the Coulomb dissociation of $^8$Li in order to extract the neutron capture cross section for the inverse reaction $^7$Li(n,$\gamma$)$^8$Li. 

A 69.5-MeV/nucleon secondary $^8$Li beam bombarded a Pb target exciting projectiles by virtual photons. Excited unbound states subsequently decay by neutron emission to $^7$Li which is stable. The neutron capture reaction on $^7$Li can be directly measured so that the validity of the Coulomb dissociation method to deduce the capture cross section of the inverse reaction can be tested. Indeed, the excitation function of the reaction $^7$Li(n,$\gamma$)$^8$Li has been measured over a wide range of energies from a few meV up to 1 MeV \cite{PhysRev.114.1037,1989ApJ...344..464W,1991ApJ...381..444N,Blackmon1996-383,
Heil1998-997,PhysRevC.71.055803}.

\begin{figure}
\includegraphics[width=1\columnwidth]{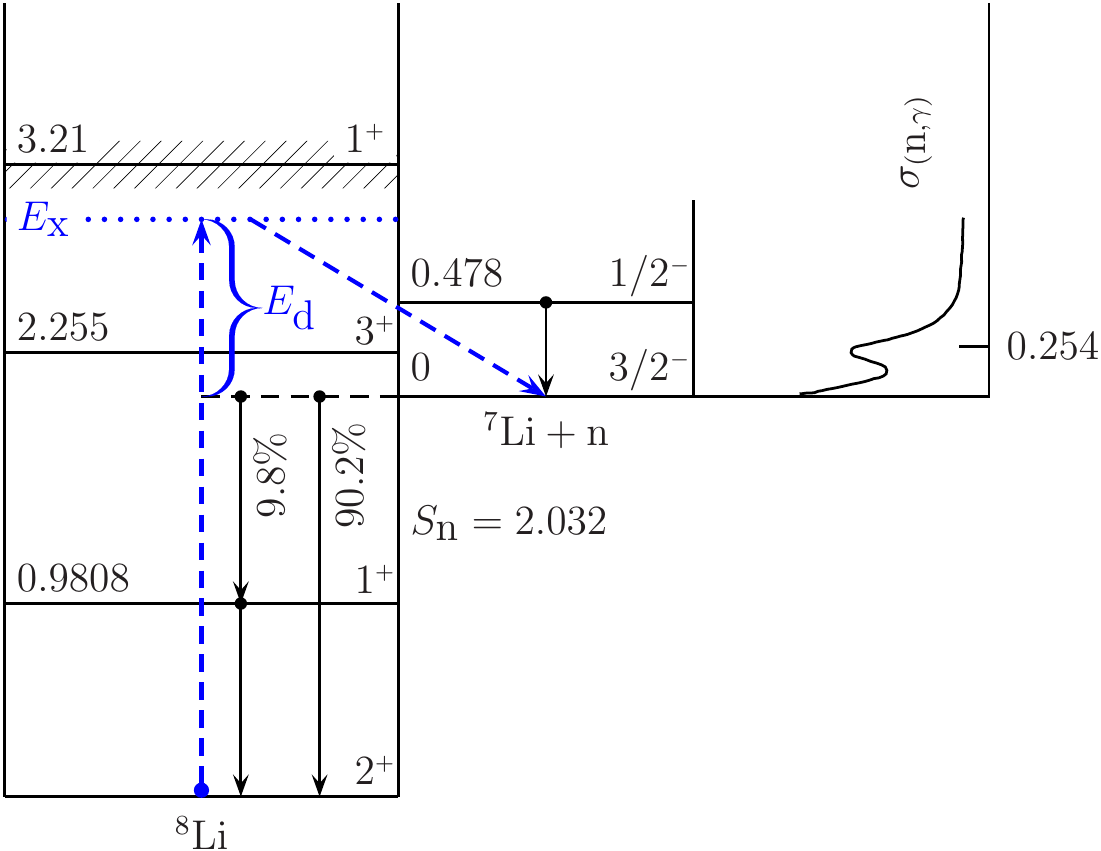}
\caption{(Color online) Partial level scheme of $^8$Li and $^7$Li. Coulomb dissociation of $^8$Li is shown by the blue dashed lines and the right panel displays the spectrum for the neutron capture reaction on $^7$Li. Energies are given in MeV. Adapted from Ref. \cite{Tilley2004155}; the $\gamma$-ray branching ratios are from Refs. \cite{EGAF2006,TUNL-nuclData}.}
\label{fig:Li-levelSchema}
\end{figure}

Partial level schemes of $^7$Li and $^8$Li are shown in Figure~\ref{fig:Li-levelSchema}. The excitation and decay during the Coulomb dissociation process of $^8$Li are indicated by the blue dashed lines. When a virtual photon from the Pb target excites the $^8$Li projectile to an excitation energy $E_\textrm{x}$ above the neutron separation energy $S_\textrm{n}$ of 2.032~MeV, $^8$Li decays to $^7$Li with a decay energy of $E_\textrm{d} = E_\textrm{x} - S_\textrm{n}$. In the direct process, a neutron is captured with energy $E_\textrm{d}$ and a $\gamma$-ray of energy $E_\textrm{x}$ is emitted. 

Before the (n,$\gamma$) cross sections derived by detailed balance from Coulomb dissociation data can be compared with the directly measured neutron capture cross section, several corrections have to be applied.  They will be discussed in detail in sections  \ref{sec:CDF-PDC} and \ref{sec:ncapture}.

\section{Experimental procedure}
\label{sec:The-experimental-procedure}

\subsection{Setup}

\begin{figure*}
\includegraphics[width=1.6\columnwidth]{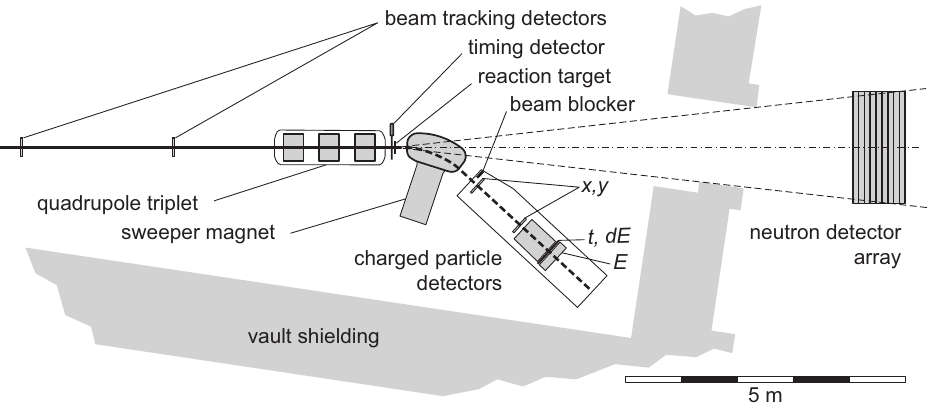}
\caption{Experimental setup. The position and angle of the incoming $^8$Li beam were measured by two beam-tracking detectors. A timing detector in front of the target served as the start for the fragment and neutron time-of-flight measurements. Neutrons were detected around 0$^\circ$ with the neutron detector array MoNA \cite{Baumann2005-517,Luther2003-33}. The charged fragments were detected by two CRDCs ($x$,$y$), a thin ($t$, $dE$), and a thick ($E$) scintillator behind a wide-gap sweeper magnet which bent the $^7$Li fragment to approximately 40$^\circ$. Unreacted $^8$Li beam particles were stopped in a beam blocker located behind the sweeper magnet at a few degrees less than 40$^\circ$ \cite{schiller:92,EPJA.27.217}.}
\label{fig:Experimental-setup}
\end{figure*}

The experiment was carried out at the National Superconducting Cyclotron Laboratory at Michigan State
University. A 120~MeV/nucleon $^{18}$O beam from the Coupled Cyclotron Facility bombarded a 2850~mg/cm$^2$ $^9$Be target. The secondary $^8$Li beam was selected by the A1900 fragment separator utilizing an 825~mg/cm$^2$ aluminum wedge degrader. The average beam intensity was $\sim$150,000/s, the mean energy was 69.5 MeV/nucleon, and the energy dispersion could be best described by a rounded rectangle with FWHM = 1.8 MeV/nucleon.  The $^8$Li particles then impinged on 56.7~mg/cm$^2$ lead and 28.8~mg/cm$^2$ carbon targets corresponding to energy losses of 2.3~MeV and 2.2~MeV, respectively. A schematic view of the experimental setup is shown in Figure \ref{fig:Experimental-setup}.
The $^8$Li beam particles were tracked with a pair of Cathode Readout Drift Chambers (CRDC) \cite{Yurkon1999-291} separated by 2.76~m through a quadrupole triplet magnet onto the reaction target. A 26.4~mg/cm$^2$ thin plastic scintillator positioned just before the target provided the start signal for time-of-flight (ToF) measurements. 

Neutrons from the breakup of $^8$Li were detected by the Modular Neutron Array
(MoNA) \cite{Baumann2005-517,Luther2003-33}. MoNA was arranged in 9 vertical layers of
16 horizontal scintillator bars each. The front face of the first layer was placed at 8.27~m from the reaction target. For the present analysis, only the first 6 layers and  the center 1.6~m of the 2-m length were used. The horizontal and vertical acceptances were $\pm$2.8\textdegree\ and $\pm$3.1\textdegree, respectively. Each bar has a photomultiplier tube mounted on each end. The time and position of an interaction in a bar are calculated by the mean value and the difference of the left and right signals, respectively. The geometric mean of the left- and right signal charges is approximately proportional to the deposited
energy. The individual bars were gain-matched using $\gamma$
rays from \textsuperscript{88}Y (1611 keV) and \textsuperscript{228}Th
(2381 keV) radioactive sources. The energy threshold was set at 0.7~MeVee. 

Charged fragments from the reaction were deflected by the sweeper
magnet \cite{Bird2005-1252} into a suite of charged-particle detectors \cite{NathanPHDthesis2006}.
Two CRDCs, separated by 1~m, determined the trajectories of
the projectile-like fragments. A thin scintillator served as the fragment trigger and provided an energy loss ($dE$) measurement. The fragments were then stopped in a thick scintillator which recorded the remaining energy ($E$). A beam blocker placed behind the sweeper magnet on the high-rigidity side stopped the unreacted $^8$Li beam in order to limit the overall count rate in the detection system. The average $^7$Li rate  entering the detectors was $\approx$ 0.3/s. 

\subsection{Incoming beam parameters}
\label{sec:Momentum-determination}

The position and angle of the incoming $^8$Li at the target are important for the determination of the overall acceptances of the $^7$Li fragments. Due to space constraints it was not possible to measure these beam parameters directly in front of the target as the target was located very closely behind a quadrupole triplet magnet (see Figure \ref{fig:Experimental-setup}). Thus, the position and angle of the incoming beam were measured event-by-event with two CRDC tracking detectors located in front of the triplet magnet. The $x$ and $y$ positions in the CRDCs were calibrated with masks to an accuracy of 0.7~mm. 
The angular straggling in the timing detector was calculated with the program {\sc Lise$^{++}$} \cite{Tarasov2008-4657} to be 0.29~mrad which was small compared to the 1.7~mrad angular spread of the fragments.  The beam trajectories through the magnet to the target were then calculated event-by-event with the particle optics code {\sc Cosy Infinity}  \cite{COSYmanual2006-21}. They were validated by bending the $^8$Li beam without a target through the sweeper magnet into the second set of CRDCs. This method has been sucecssfully used in several previous experiments \cite{PhysRevC.78.044303,PhysRevC.85.034327,Mosby201369}.

\subsection{Reconstruction of neutron energy and momentum vector}

In order to reconstruct the decay energy spectrum the energy and momentum vector of the neutrons had to be reconstructed. The neutron direction was deduced from the position of the interaction in a MoNA scintillator bar relative to the target. The position resolution was 7~cm FWHM in the horizontal direction and 10~cm full width in the vertical direction determined by the height of the bars. The magnitude of the momentum as well as of the energy were calculated from the flight time and the path length. The absolute time for the central bars of each layer was calibrated using prompt $\gamma$ rays from the target. The other bars within each layer were then synchronized to the central bar with cosmic rays. A time resolution of 1.15~ns (FWHM) was achieved which was dominated by the uncertainty of the flight path due to the 10~cm thickness of the bars.

\begin{figure}
\includegraphics[width=1\columnwidth]{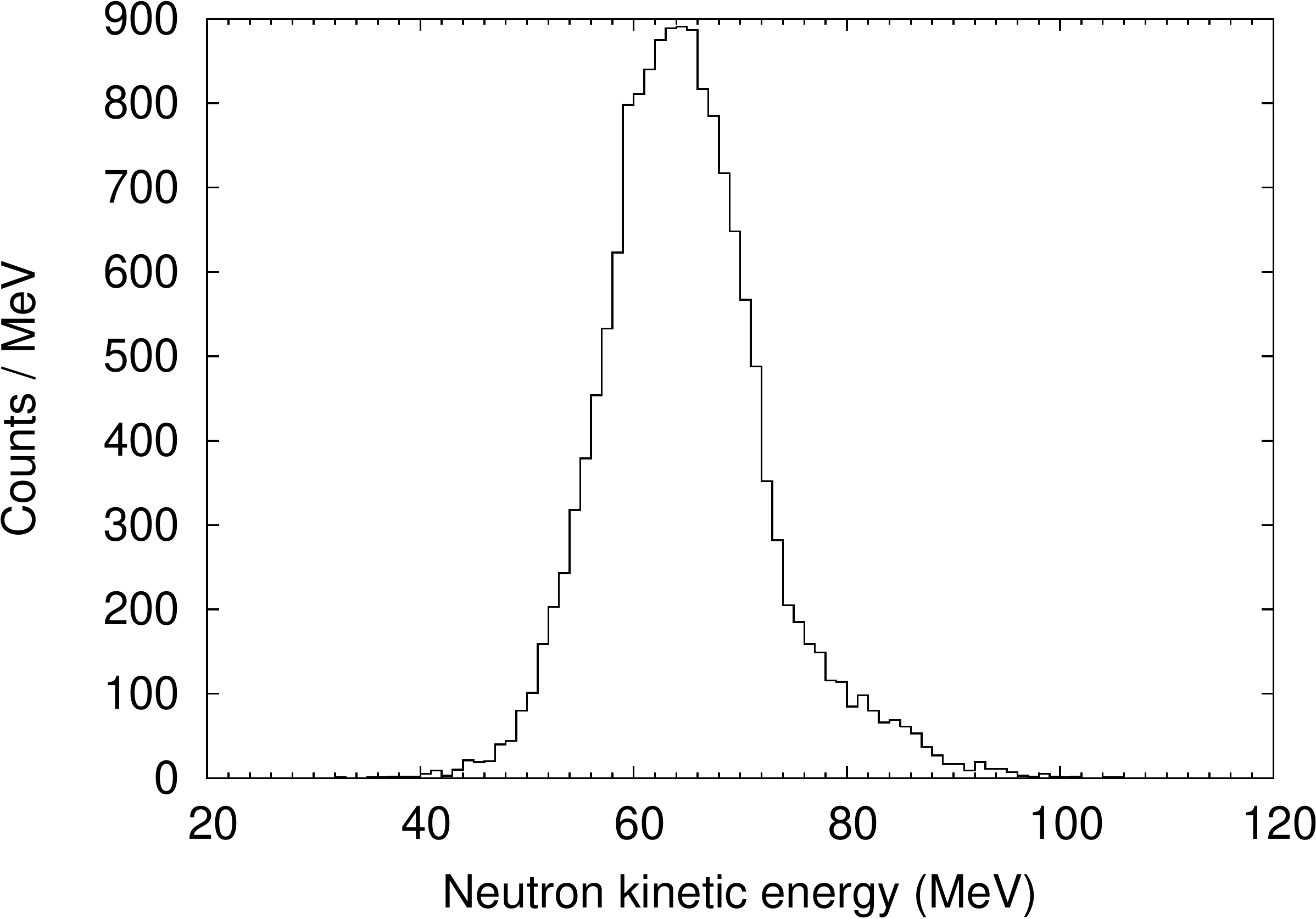}
\caption{Neutron energy spectrum gated on events with a decay energy of $E_\textrm{d} < 1.5$~MeV.}
\label{fig:Neutron-energy-spectrum}
\end{figure}

The neutron energy spectrum for the events of interest ($E_\text{d} < 1.5$~MeV)\footnote{The decay energy ($E_\text{d}$) is calculated using Eq. (2); only events with $E_\text{d} < 1.5$~MeV were included in the further analysis.} is shown in  Figure \ref{fig:Neutron-energy-spectrum}. The distribution peaks around 64 MeV which is about 5.5~MeV below the energy per nucleon of the $^8$Li beam. The energy loss in the target and the binding energy of the neutron in $^8$Li together account for only 0.5~MeV/nucleon. Most of the reduction is due to the increase of the Coulomb potential energy of the beam as it approaches the Pb target nuclei. The neutron is emitted near the distance of closest approach where the kinetic energy is reduced by about 38~MeV or close to 5~MeV/nucleon.

The apparent asymmetry of the peak arises from the fact that relative to the beam velocity forward emitted neutrons gain more energy than backward emitted neutrons lose energy. In addition the angular acceptance for forward emitted neutrons is larger than for backward emitted neutrons.

\subsection{Reconstruction and identification of $^7$Li}
\label{sub:Li-fragments-sweeperM}

\begin{figure}
\includegraphics[width=1\columnwidth]{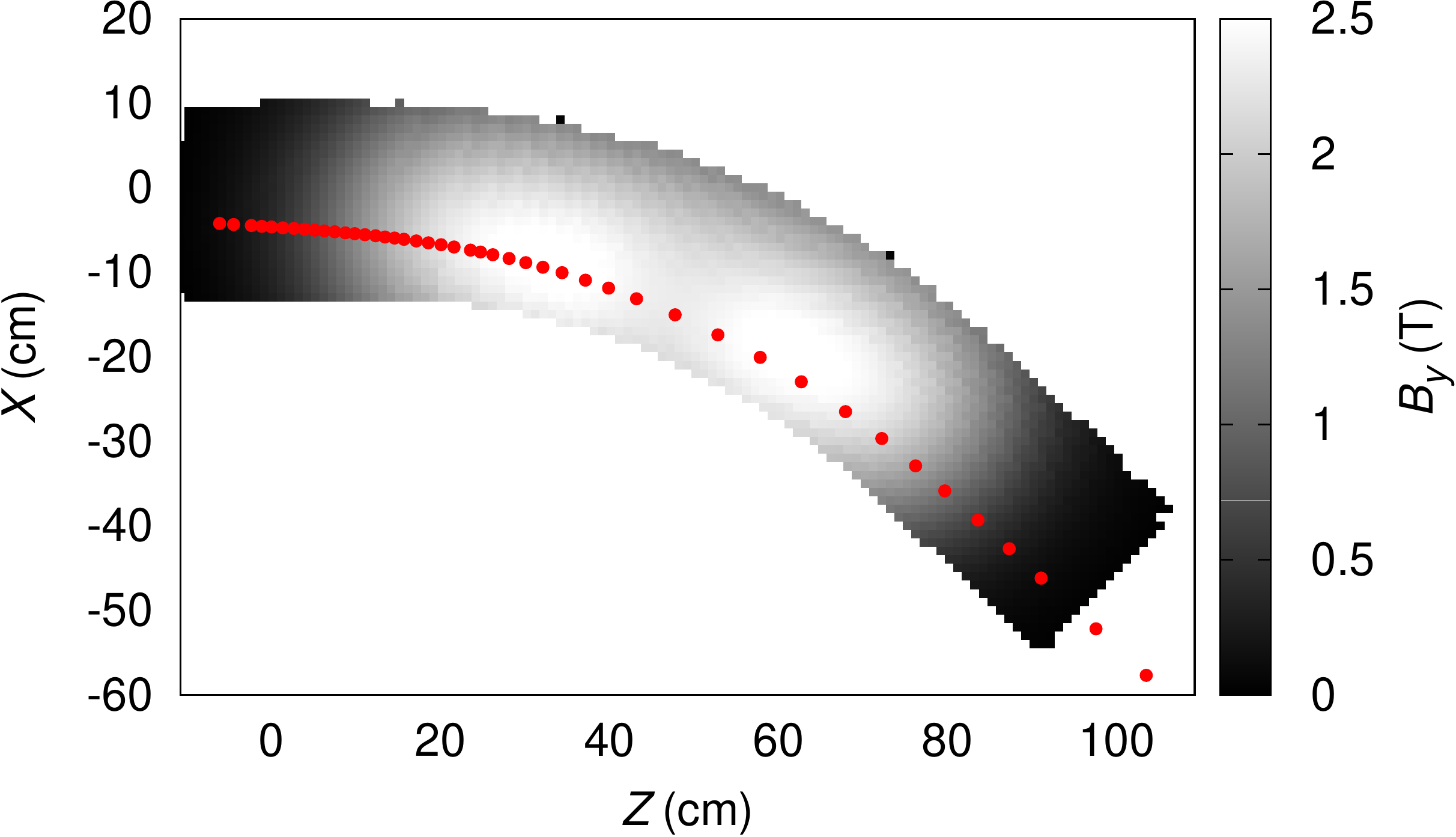}
\caption{(Color online) Sweeper magnet field map. The direction of the incident beam is in the positive Z-direction and horizontal deflection of the magnet is in the negative X-direction. A typical calculated trajectory through the magnet is shown by the red circles.}
\label{fig:The-field-map}
\end{figure}

The momenta of the $^7$Li fragments were calculated from the tracks through the magnetic field of the sweeper magnet and the positions measured in the CRDCs.
A field map of the sweeper magnet had been generated in seven horizontal layers with a grid of 6,000 points each \cite{NathanPHDthesis2006}.  Within a horizontal layer the magnetic field at any point was linearly extrapolated from a triangular mesh. In the vertical direction, the field was determined with a spline interpolation from seven values along a vertical line
crossing the layers. The fringe field was extrapolated with Enge functions \cite{COSYmanual2006-21}. During the experiment, the field was monitored with a Hall probe. An adaptive fifth-order Runge-Kutta method \cite{numRec1992} was used to track the fragments through the magnetic field, solving the equation of motion numerically.
Starting with the known trajectory after the magnet as measured by the positions in the two CRDCs, particles with different trial momenta were backtracked through the magnet until their position coordinates at the target matched the reaction point determined by the incoming particle (see Section \ref{sec:Momentum-determination}). The end point of the matching
trajectories yielded the momentum vector of the $^7$Li fragments at the reaction point. The Sweeper Magnet field map and a typical particle trajectory is shown in Figure \ref{fig:The-field-map}. This method of momentum reconstruction was validated by bending the $^8$Li beam without a target through the sweeper magnet into the CRDCs. The reconstructed momenta could then directly be compared with the momenta of the incoming particles.

\begin{figure}
\includegraphics[width=1\columnwidth]{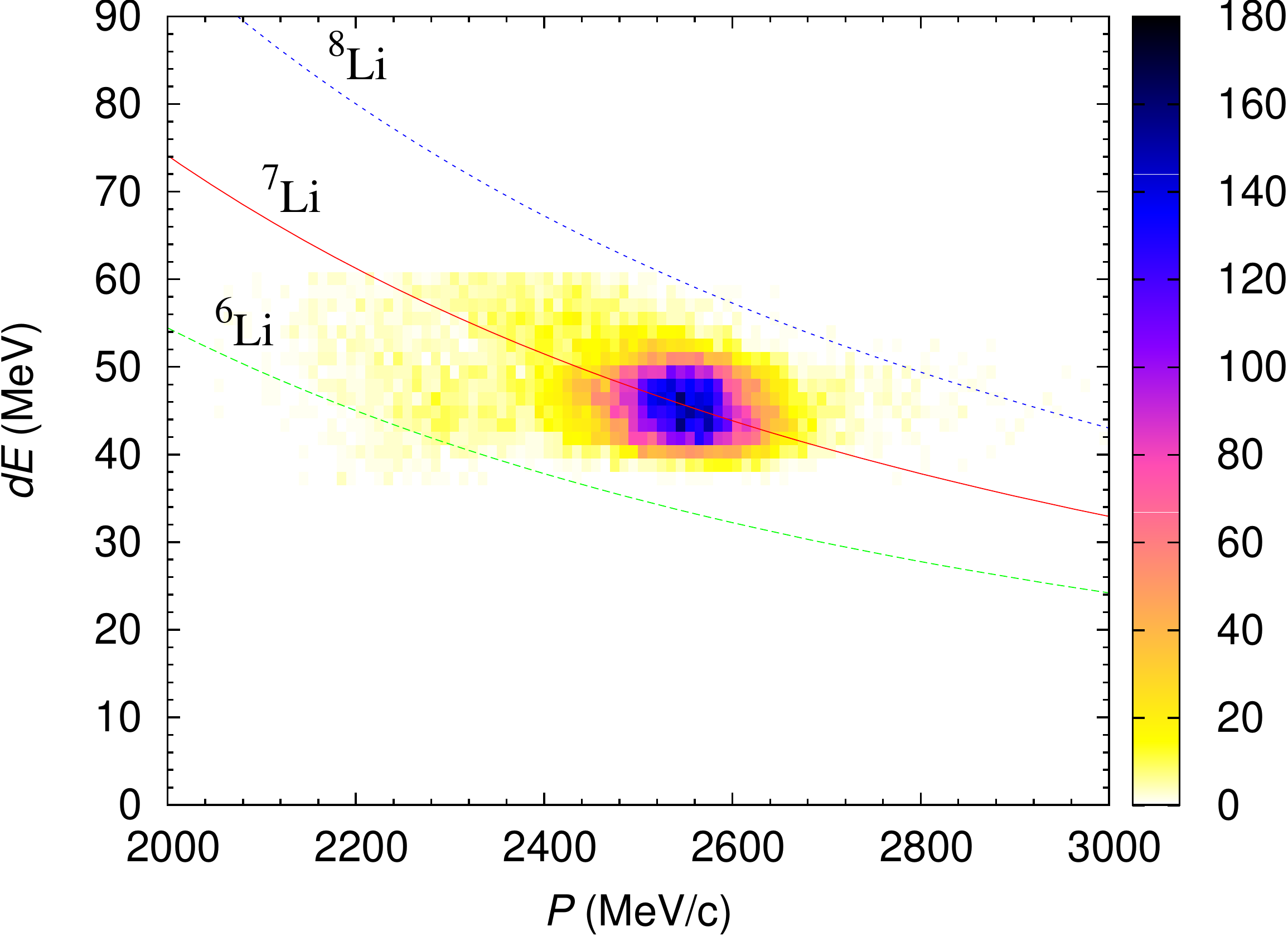}
\caption{(Color online) Energyloss in the thin scintillator versus momentum. The $^7$Li fragments are cleanly separated from some minor contributions of $^6$Li fragments.}
\label{fig:ID}
\end{figure}

\begin{figure}
\includegraphics[width=1\columnwidth]{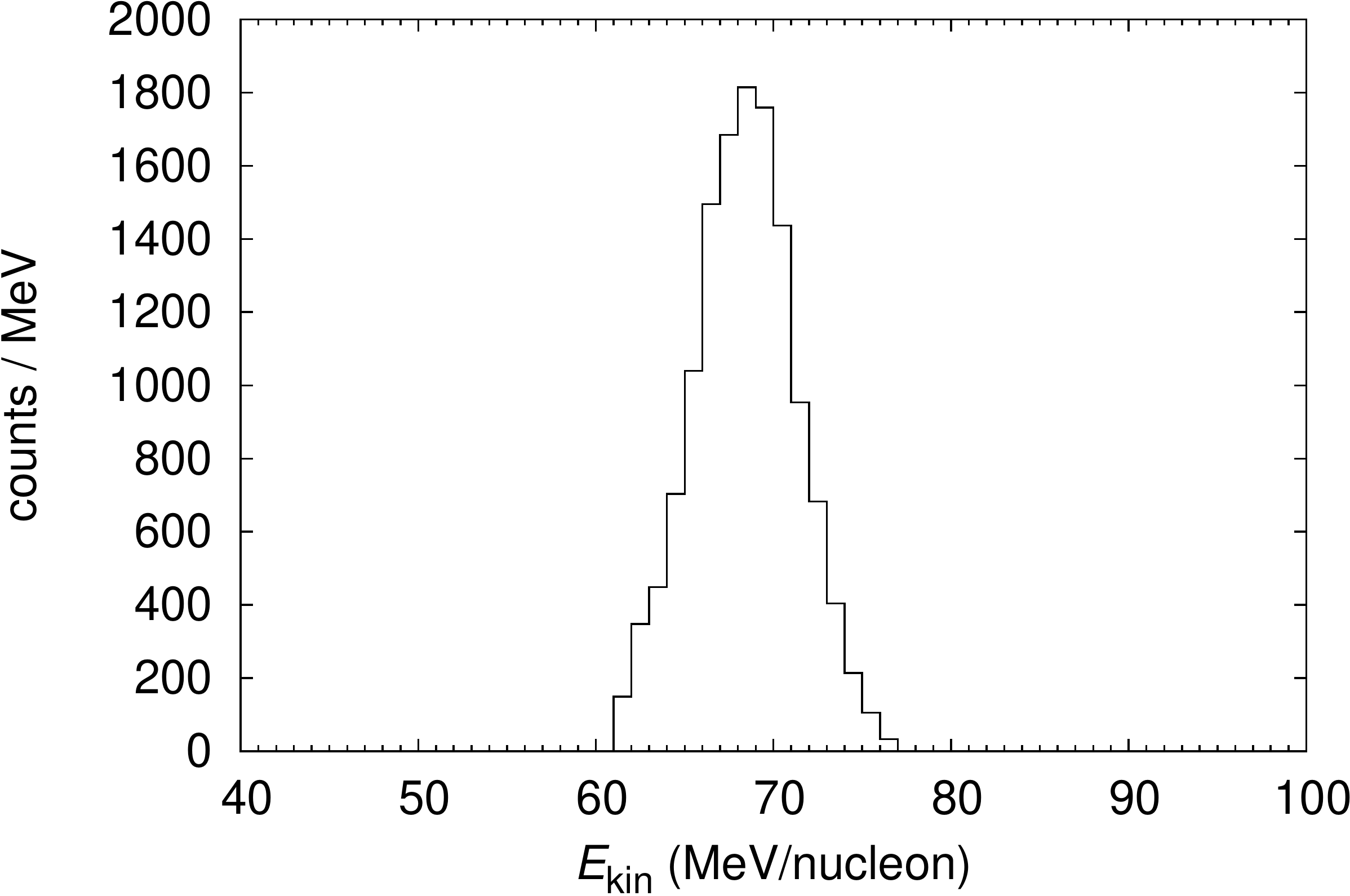}
\caption{$^7$Li energy spectrum gated on events with a decay energy of $E_\textrm{d} < 1.5$~MeV.}
\label{fig:Li-energy-spectrum}
\end{figure}

Lithium fragments were identified by the energy deposited while passing through the thin scintillator located after the CRDCs. Isotopic separation was achieved from a two-dimensional spectrum of energy loss versus momentum as shown in Figure \ref{fig:ID}. The energy spectrum of $^7$Li fragments for events with $E_\textrm{d} < 1.5$~MeV is shown in Figure \ref{fig:Li-energy-spectrum}. In contrast to the neutron energy spectrum shown in Figure \ref{fig:Neutron-energy-spectrum} the $^7$Li distribution peaks close to the energy per nucleon of the incoming $^8$Li beam. As mentioned earlier, the reduction due to the neutron binding energy and the energy loss in the target is small and the potential energy lost by the $^8$Li as it approaches the Pb nuclei is regained by the $^7$Li fragment after the breakup. The fragment energy distribution is also symmetric and narrower than the neutron spectrum which is due to the smaller velocity imparted to the fragment in the breakup.

\section{Data analysis}
\label{sec:Analysis}

\subsection{Decay energy spectrum}

The decay energy ($E_\textrm{d}$) from the breakup of $^8$Li can be calculated from the energy and momenta of the neutrons ($E_\textrm{n}$, $\vec{p}_\textrm{n}$) and $^7$Li fragments ($E_{^7\textrm{Li}}, \vec{p}_{^7\textrm{Li}}$) as
\begin{equation}
E_\textrm{d} = \sqrt{(E_\textrm{n} + E_{^7\textrm{Li}})^2 - |\vec p_\textrm{n} + \vec p_{^7\textrm{Li}}|^2} - m_\textrm{n} - m_{^7\textrm{Li}}
\end{equation}
where $m_\textrm{n}$ and  $m_{^7\textrm{Li}}$ are the rest masses of the neutron and $^7$Li, respectively. Before the decay energy spectrum can be converted to the Coulomb dissociation function and subsequently the photo disintegration cross section, the data have to be corrected for the efficiencies and acceptances of the detector systems.

The intrinsic efficiency of MoNA was simulated with {\sc Geant4} \cite{Agostinelli2003250,GEANT4-2006} using the {\sc Menate\_R} package \cite{Roeder2008} as described in \cite{Kohley201259}. In addition to the properties of MoNA itself, the simulations included the window of the vacuum chamber (1/4-inch of stainless steel) and the 827~cm of air between the target and the front face of MoNA. The efficiency varied linearly for neutron energies of interest from 75.3(30)\% at 50~MeV to 66.1(26)\% at~80 MeV. For the Coulomb dissociation function below 1.5~MeV the efficiency did not vary as function the neutron energy so that the overall spectrum could be corrected by 70.5(28)\% corresponding to the efficiency at the average neutron energy of 65~MeV.

The efficiencies of the charged particle detectors were determined by removing the target and bending the $^8$Li beam into the center of the focal plane detectors. The combined efficiency for the two tracking CRDC detectors (84.4(8)\%), the target timing scintillator (99.9(10)\%), the focal plane CRDC detectors (95.0(10)\%), $dE$ and $E$ plastic scintillators (99.7(10)\% each) was 79.0(16)\%.

\begin{figure}
\includegraphics[width=1\columnwidth]{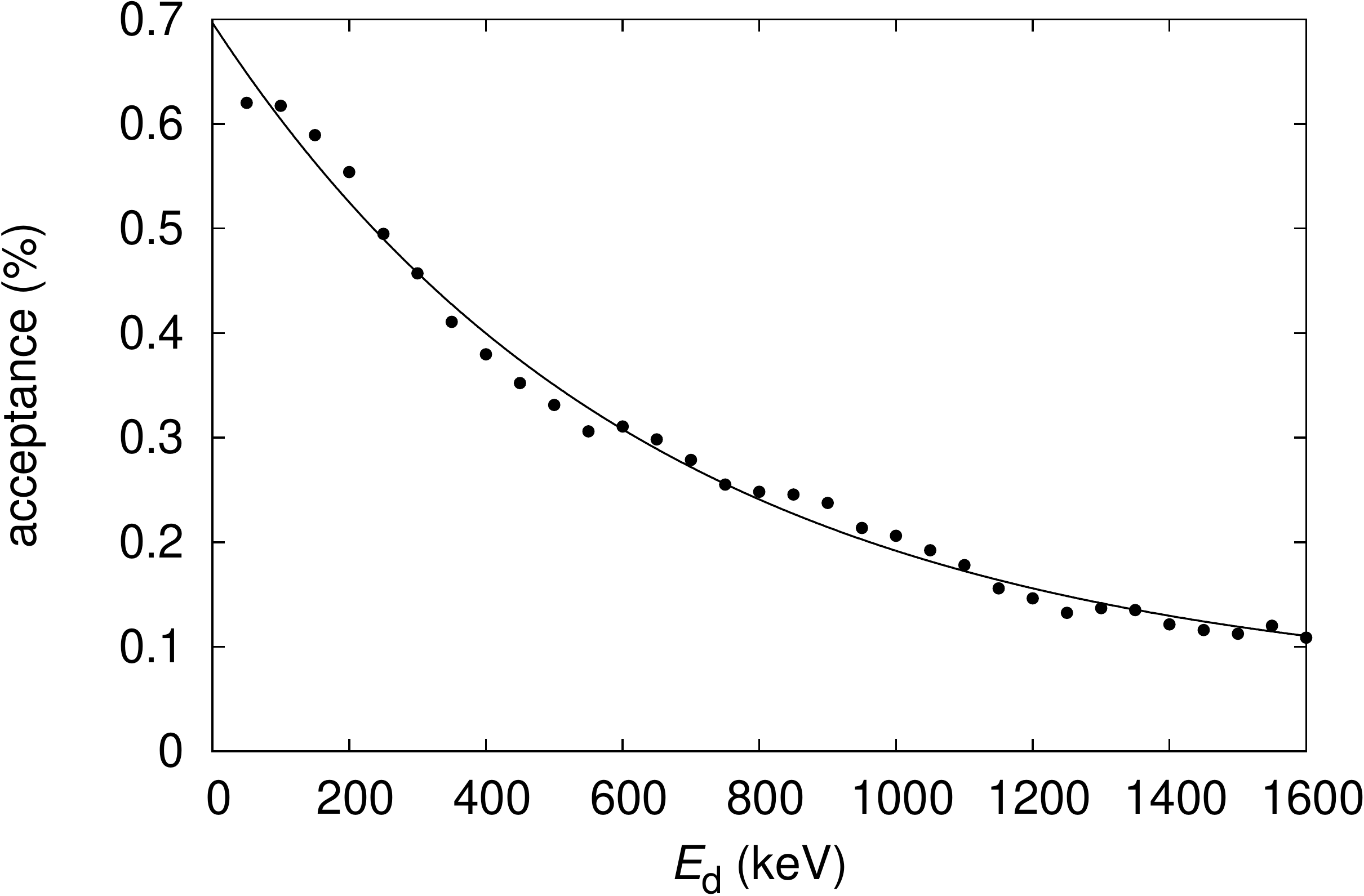}
\caption{Overall solid-angle acceptance of the neutron and $^7$Li detectors. The circles correspond to the results of the Monte Carlo simulation and the dashed line represents a fit as described in the text.}
\label{fig:MonteCarlo-acceptance}
\end{figure}

The overall acceptance for the detection of the neutrons and $^7$Li fragments is correlated and depends on the decay energy. Thus, a Monte Carlo event-by-event simulation was written  which included the properties of the incoming beam, the reaction mechanism and the geometry of the detectors including the beam blocker. The trajectories of $^7$Li fragments were simulated following Rutherford scattering, and the energy of the virtual photon was selected according to the description by Baur and Bertulani \cite{Bertulani1985-739}. The breakup into a neutron and $^7$Li was assumed to be isotropic in the rest frame of $^8$Li. The results of the simulation are shown as the circles in Figure \ref{fig:MonteCarlo-acceptance}. The dashed line corresponds to a fit of the form $(1-c)/(e^{a(E_\textrm{d}-b)})+c$ which was used to correct the decay energy spectrum for the detector efficiencies and acceptances.

The uncertainty of this correction was calculated to be 4.5\% by combining the solid angle uncertainties of MoNA and the uncertainties of the fragment detector system which were estimated to be 4\% and 2\%, respectively.

\subsection{Coulomb dissociation function and photo disintegration cross section}
\label{sec:CDF-PDC}

The decay energy spectrum contains contributions from Coulomb dissociation as well as nuclear reactions. Peripheral Coulomb breakup increases with the charge of the projectile and the target while the cross section for peripheral breakup due to the short-range nuclear force depends on the radii of the projectile and target nuclei. In order to subtract the nuclear contribution from the decay energy spectrum, the breakup spectrum was measured with a carbon target. The relative contributions from Coulomb and nuclear reactions were then parameterized as a function of the mass number ($A$) and charge ($Z$) of the target as described in Ref. \cite{Horvath2002-926}:
\begin{equation}
\frac{d\sigma(E_\textrm{d})}{dE_\textrm{d}} = a(E_\textrm{d})(r_0A^{1/3} + r_{^8\textrm{Li}}) + b(E_\textrm{d}) Z^{1.85}
\end{equation}
with $r_0$ = 1.2~fm and $r_{^8\textrm{Li}}$ = 2.4~fm. The parameters a($E_\textrm{d}$) and b($E_\textrm{d}$) were determined from the measured cross sections of the lead and carbon targets and were about 0.6~mb/(MeVfm) and 0.03~mb/MeV, respectively. The nuclear contribution to the decay energy spectrum for the lead target was between 3\% and 5\%
for $E_\textrm{d}$ up to 1.5 MeV. Coulomb-nuclear interference effects were determined by a DWBA calculation to be more than two orders of magnitude smaller than the individual contributions. The above paramerization is very similar to the approach by Fukuda {\it et al.} \cite{PhysRevC.70.054606} who measured the breakup cross sections of $^{11}$Be on Pb and C targets and deduced the Coulomb dissociation cross section according to $\sigma_{\textrm{CD}} = \sigma(Pb) - \Gamma \sigma(C)$. The scaling factor $\Gamma$ of 2.1(5) was extracted from angular distribution measurements. Converting our paramerization to $\Gamma$ yields a value of $\sim$1.5. The angular acceptance was the same ($<6^\circ$) in both experiments.

It should also be mentioned that only the relative decay energy of each neutron-$^7$Li coincident pair is determined and a given decay energy can thus result from decay to either the 3/2$^-$ ground state or the 1/2$^-$ first excited state. However, using a continuum coupled-channels calculation with states generated with a potential model \cite{PhysRevLett.94.072701} transitions to the excited state by Coulomb dissociation were determined to be more than three orders of magnitude smaller than transitions to the ground state; thus these contributions are considered negligible.

\begin{figure}
\includegraphics[width=1\columnwidth]{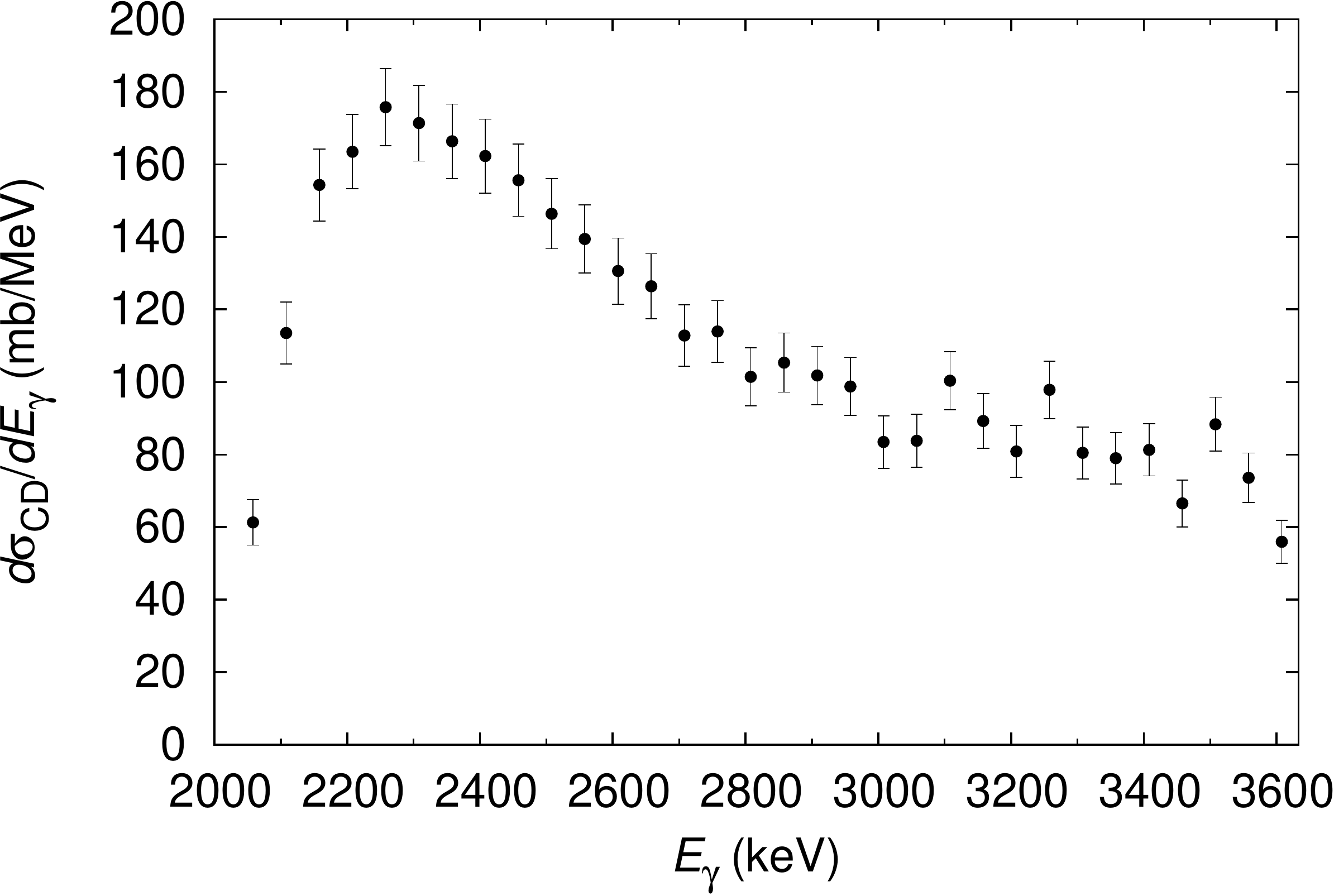}
\caption{Measured Coulomb dissociation function for the breakup of $^8$Li.}
\label{fig:Measured-Coulomb-excitation}
\end{figure}

The Coulomb dissociation function, which is directly related to the corrected decay energy spectrum by the substitution $E_{\gamma} = E_\textrm{d} + 2.032$~MeV, is shown in Figure \ref{fig:Measured-Coulomb-excitation}. The photo disintegration cross section ($\sigma_{\gamma,\textrm{n}}$) can then be calculated by the virtual photon method according to Eq. (\ref{eq:virtual-photon-method}).

\subsection{Neutron capture cross section}
\label{sec:ncapture}

Finally, the neutron capture cross section for the direct reaction $^7$Li(n,$\gamma$)$^8$Li is derived by the principle of detailed balance given by
\begin{equation}
\sigma_{\textrm{n},\gamma_0}=\frac{E_{\gamma}^{2}}{2\mu E_{\textrm{c.m.}}c^{2}}\frac{2(2j_{^8\textrm{Li}}+1)}{(2j_{^7\textrm{Li}}+1)(2j_\textrm{n}+1)}\sigma_{\gamma_0,\textrm{n}}\label{eq:Detailed-balance}
\end{equation}
where $\mu$ is the reduced mass and $E_{\textrm{c.m.}}$ is the energy in the center-of-mass system in the n + $^7$Li capture process which
is equivalent to $E_\textrm{d}$ in the $\gamma$-ray dissociation reaction. This cross section only corresponds to the neutron capture to the ground state of $^8$Li ($\sigma_{\textrm{n},\gamma_0}$). However, the direct capture reaction can also proceed via the bound first excited state ($\sigma_{\textrm{n},\gamma_1}$). Thus, the Coulomb dissociation data have to be corrected for these contributions. The total cross section can be expressed as:
\begin{equation}
\sigma_{\textrm{n},\gamma} = \sigma_{\textrm{n},\gamma_0} \left(1+ \frac{\sigma_{\textrm{n},\gamma_1}}{\sigma_{\textrm{n},\gamma_0}}\right).
\label{eq:crossSecRatios}
\end{equation}
The ratio {$\sigma_{\textrm{n},\gamma_1}$}/{$\sigma_{\textrm{n},\gamma_0}$} can be evaluated at thermal neutron energies ($E_\textrm{d} = 0$) where the branching ratio is known and because both $\gamma$-ray decays are E1 transitions, the relative intensities scale as $E_{\gamma}^{3}$:
\begin{equation}
\sigma_{\textrm{n},\gamma} =\sigma_{\textrm{n},\gamma_0} \left[1+\frac{BR_1}{BR_0}\left(\frac{E_{\gamma_1}+E_{\textrm{d}}}{E_{\gamma_0}+E_{\textrm{d}}}\right)^{3}\right].
\label{eq:crossSecngamma}
\end{equation}
The branching ratios ($BR$) and $\gamma$-ray energies for thermal capture ($E_\gamma$) are 9.800(344)\% and 90.20(260)\% \cite{EGAF2006,TUNL-nuclData} and 1.051~MeV ($S_\textrm{n} - E^*_1$, see Figure \ref{fig:Li-levelSchema}) and 2.032~MeV for the first excited and ground state, respectively.
The correction factor $\sigma_{\textrm{n},\gamma}$/$\sigma_{\textrm{n},\gamma_0}$
increases from 1.109(5) at $E_\textrm{d}$ = 0 to 1.243(13) at $E_\textrm{d}$ = 1.0 MeV.

\begin{figure}
\includegraphics[width=1\columnwidth]{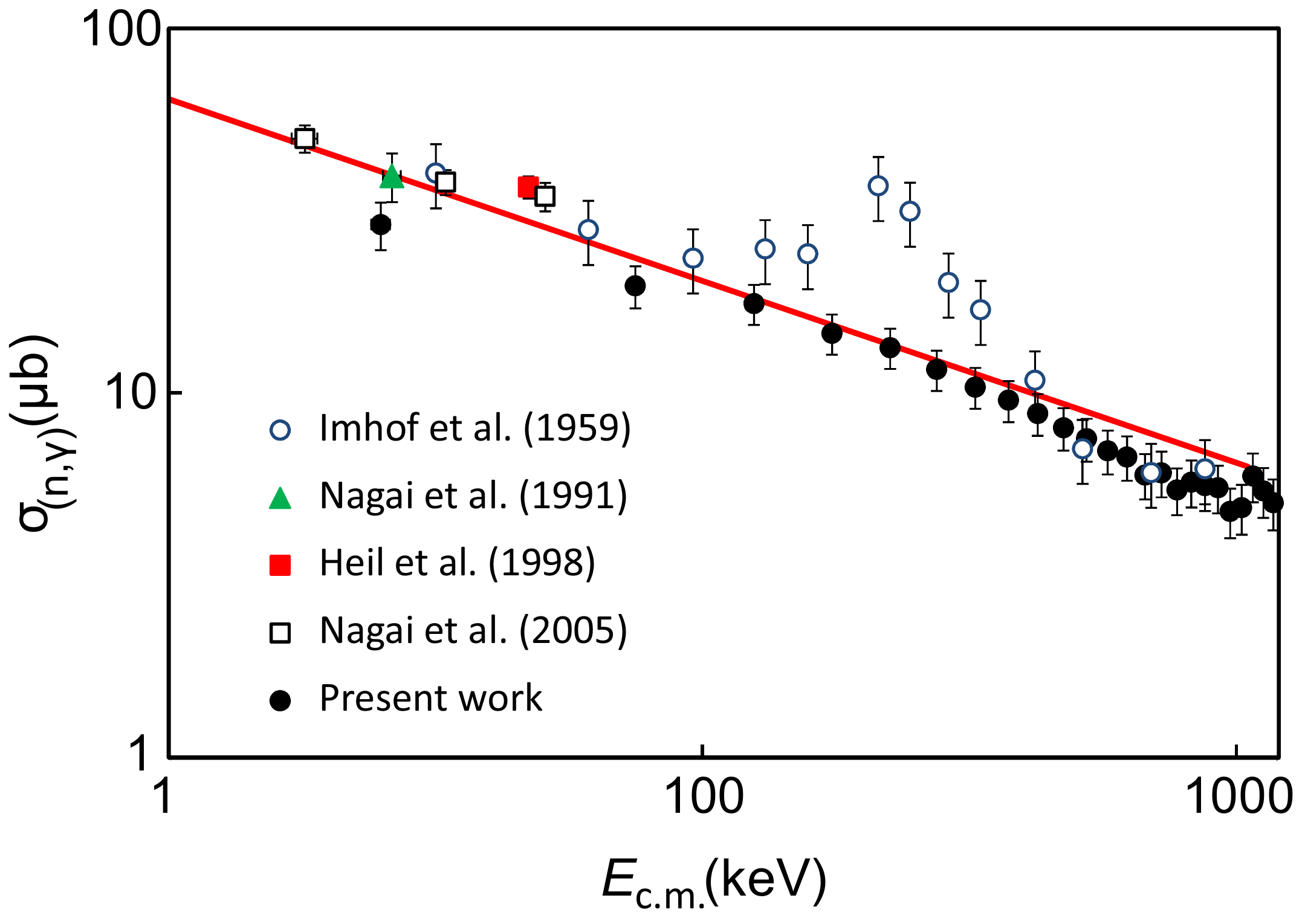}
\caption{(Color online) $^7$Li(n,$\gamma$)$^8$Li excitation functions. The data from the present Coulomb dissociation measurement (solid black circles) are compared to previous direct measurements by Imhof {\it et al.}  (open blue circles, \cite{PhysRev.114.1037}), Nagai {\it et al.} (solid green triangle, \cite{1991ApJ...381..444N}), Heil {\it et al.} (solid red square, \cite{Heil1998-997}), and Nagai {\it et al.} (open black squares, \cite{PhysRevC.71.055803}). The red solid line corresponds to the inverse velocity (1/$v$) dependence of the cross section which fit the low-energy data of Blackmon {\it et al.}  \cite{Blackmon1996-383}.}
\label{fig:Comparison-of-neutron-capture-exp}
\end{figure}

Another correction results from the fact that $s$-wave neutron capture into 1$^+$ and 2$^+$ states results in $\gamma$ rays of only E1 multipolarity while Coulomb dissociation can be induced by both E1 and E2 photons. For a quantitative estimate, we calculated $\sigma_{\gamma,\textrm{n}}$ with a modified version of RADCAP \cite{Bertulani2003-123} and found that E2 transition strengths are at least a factor 10$^6$ smaller than E1 transition strengths. 
In addition, as shown in the right panel of Figure \ref{fig:Li-levelSchema}, the direct capture reaction can proceed via the 3$^+$ resonance of $^8$Li. In Coulomb dissociation, this resonance would have to be excited by M1 transitions from the 2$^+$ ground state of $^8$Li. However, the number of virtual M1 photons is negligibly small. 
Thus, to better than a few percent, the Coulomb dissociation measures only the direct component of the neutron capture reaction. 

The final neutron capture cross section as extracted from the inverse Coulomb dissociation measurement is shown by the solid black circles in Figure \ref{fig:Comparison-of-neutron-capture-exp}. The data are compared with previous direct neutron capture measurements by Imhof {\it et al.} \cite{PhysRev.114.1037} (open blue circles), Nagai {\it et al.} \cite{1991ApJ...381..444N} (solid green triangle), Heil {\it et al.} \cite{Heil1998-997} (solid red square), and another more recent measurement by Nagai {\it et al.} \cite{PhysRevC.71.055803} (open black squares). The 254-keV resonance corresponding to the 3$^+$ second excited state in $^8$Li is clearly visible in the direct capture data by Imhof {\it et al.} but absent (as expected) in the present Coulomb dissociation data. Overall, the non-resonant data from the two different methods are in good agreement; while the present data essentially agree with the Imhof data above the 254-keV resonance (within less than 10\%), at energies below the resonance they are  systematically about 20-25\% lower than the previous data, although still within the 15-20\% uncertainties of the individual data points. At energies above $\sim$200~keV the data deviate from the expected inverse velocity (1/$v$) dependence for $s$-wave capture (red solid line) which has been well established at lower energies \cite{Blackmon1996-383} as will be discussed in the next section.

\section{Comparison to Theory}

\begin{figure}
\includegraphics[width=1\columnwidth]{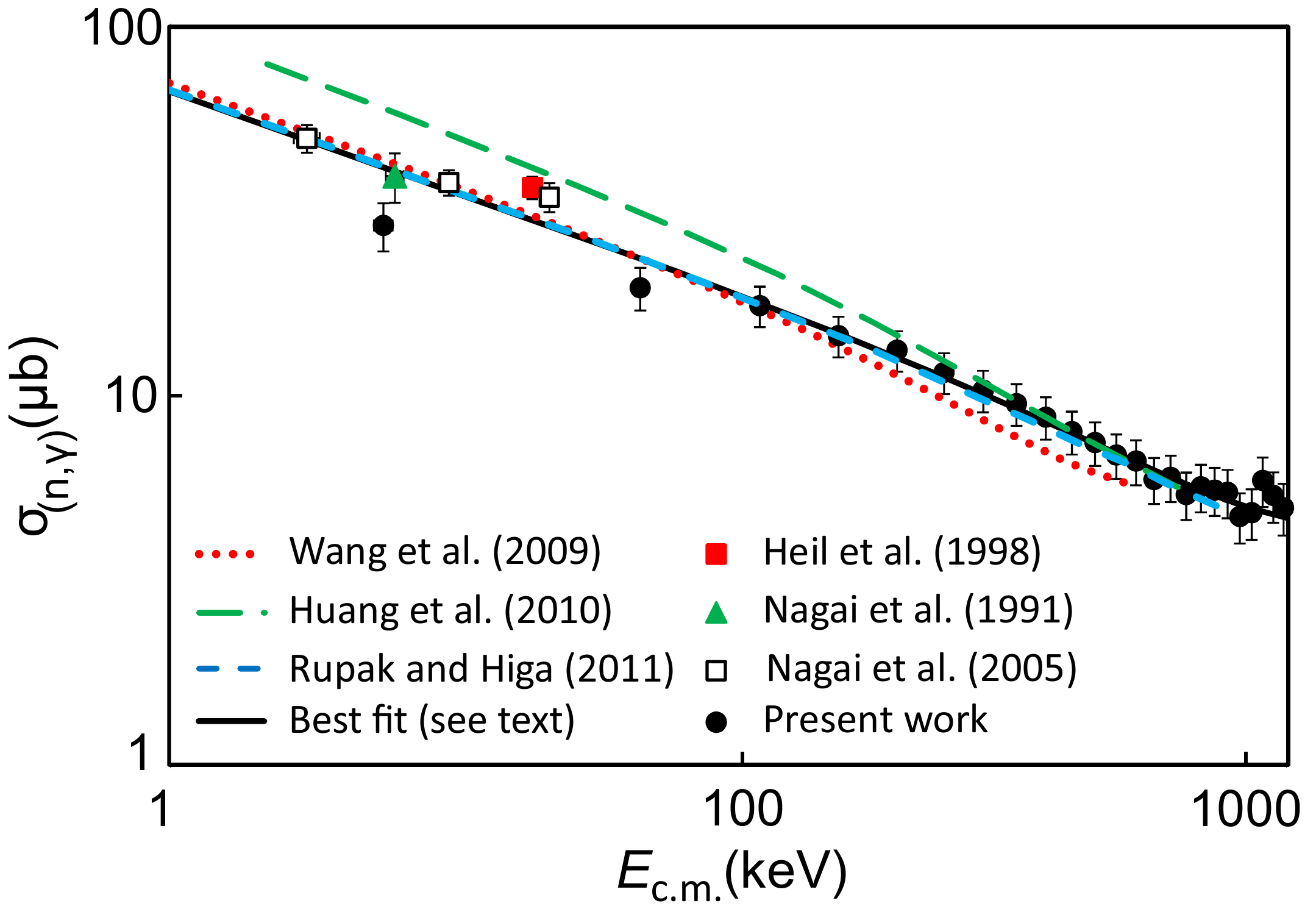}
\caption{(Color online) $^7$Li(n,$\gamma$)$^8$Li excitation functions. The non-resonant data are the same as in Figure \ref{fig:Comparison-of-neutron-capture-exp}. They are compared to calculations by Wang {\it et al.} (red dotted line, \cite{PhysRevC.80.034611}), Huang {\it et al.} (green long-dashed line, \cite{Huang2010824a}), and Rupak and Higa (blue short-dashed line, \cite{PhysRevLett.106.222501}). The best fit to the data is shown by the black solid curve.}
\label{fig:Comparison-of-neutron-capture-theory}
\end{figure}

In addition to the models described in the experimental papers \cite{1989ApJ...344..464W,Heil1998-997,PhysRevC.71.055803}, many theoretical calculations have described the available data (see for example  \cite{Descouvemont1994341, Bennaceur1999289,PhysRevC.73.024612,PhysRevC.80.034611,Huang2010824a,PhysRevLett.106.222501,IJMPE.22.1350028,
PAN-Dubovichenko}).

In Figure \ref{fig:Comparison-of-neutron-capture-theory} the non-resonant neutron-capture data are compared to some of the most recent calculations by Wang {\it et al.} (red dotted line, \cite{PhysRevC.80.034611}), Huang {\it et al.} (green long-dashed line, \cite{Huang2010824a}), and Rupak and Higa (blue short-dashed line, \cite{PhysRevLett.106.222501}). All models predict the observed deviation from the 1/v dependence of the cross section towards higher energies. The potential models for direct radiative capture by Nagai {\it et al.} \cite{PhysRevC.71.055803} and Wang {\it et al.} \cite{PhysRevC.80.034611} slightly overpredict the magnitude of the deviation. Huang {\it et al.} described the radiative proton- and neutron-capture for all available data on light nuclei with a simplified two-body treatment of the capture process \cite{Huang2010824a}. While their fit describes the present data above $\sim$200~keV, it overpredicts the available data at lower energies. The best description of the data over the whole energy range is given by the recent low-energy effective field theory calculation by Rupak and Higa \cite{PhysRevLett.106.222501}. Their calculation also fits the available data in the eV range \cite{Blackmon1996-383} and at thermal energies \cite{PhysRevC.44.764} which are not shown in Figure \ref{fig:Comparison-of-neutron-capture-theory}.

The deviation from the 1/$v$ dependence has previously been included in the parameterization of the cross section as \cite{PhysRevC.70.015801,PhysRevC.80.034611}
\begin{equation}
\sigma = s_0(1 + s_1E + s_2 E^2)/E^{1/2}
\end{equation}
The solid black line in Figure \ref{fig:Comparison-of-neutron-capture-theory} shows the best fit to the available non-resonant data in the range from 10 -- 1000~keV and where $s_0$ was fixed at $6.7~\mu$b(MeV)$^{1/2}$ to fit the measured cross section at thermal neutron energies \cite{PhysRevC.44.764}. The extracted values for $s_1 = -0.53(44)$~MeV$^{-1}$ and $s_2 = 0.3(6)$~MeV$^{-2}$ are not well constrained. If the fit is not constrained at the thermal data point it yields values of $s_0 = 6.4(3)~\mu$b(MeV)$^{1/2}$, $s_1 = -0.16(61)$~MeV$^{-1}$ and $s_2 = -0.16(81)$~MeV$^{-2}$ where $s_1$ and $s_2$ are even more uncertain. In either case the values for $s_1$ and $s_2$ are smaller than the values extracted from the semiempirical parameterization of Wang {\it et al.\ }~\cite{PhysRevC.80.034611} of $-1.37$~MeV$^{-1}$ and 1.25~MeV$^{-2}$, respectively. We note that the large value of $s_2$ in this parameterization results in a positive curvature and as mentioned by Wang {\it et al.} is not appropriate above 600~keV. In order to describe the data at higher energy a higher-order polynomial description might be necessary.
The value for $s_0$ from the unconstrained fit can be compared to the slope extracted from the direct measurement of $\sigma_{\textrm{n},\gamma}$ for energies between $\sim$1~eV and $\sim$1~keV by Blackmon {\it et al.}~\cite{Blackmon1996-383}. After transforming to the laboratory system and correcting for the excited state contribution it translates to a slope of  6.2(3)$\times 10^{-3}$~b(eV)$^{1/2}$ which agrees well with the value of 6.3(3)$\times 10^{-3}$~b(eV)$^{1/2}$ quoted by Blackmon {\it et al.}.

\section{Conclusion}

In conclusion, the Coulomb dissociation reaction $^8$Li($\gamma$,n)$^7$Li was measured with a 69.5 MeV/nucleon $^8$Li beam on a Pb target, and the decay energy spectrum was calculated by measuring the outgoing neutron and $^7$Li nucleus in coincidence.
From these data and the principle of detailed balance the neutron capture cross section $^7$Li(n,$\gamma$)$^8$Li was deduced for energies between 25~keV and 1.5~MeV. The good agreement with directly measured cross sections demonstrates that Coulomb dissociation is a reliable method to extract neutron capture cross sections. It represents the first time that neutron capture cross sections for $s$-wave neutrons were derived from Coulomb dissociation. The only other case where the ($\gamma$,n) Coulomb dissociation cross sections were compared with directly measured cross sections was the system $^{14}$C(n,$\gamma$)$^{15}$C which involved predominantly the capture of $p$-wave neutrons \cite{Nakamura2009-035805}.
The anticipated deviation from the 1/v behavior at higher neutron energies was observed and could be fitted within the parameterization of Baye \cite{PhysRevC.70.015801} and Wang {\it et al.}  \cite{PhysRevC.80.034611}. The data also agree well with the results of the low-energy effective field theory calculations by Rupak and Higa \cite{PhysRevLett.106.222501}.

\section*{Acknowledgments}

Support of the National Science Foundation under grant Nos. PHY01-10253,
PHY03-54920, PHY04-56463, PHY06-06007, PHY11-02511, the Department of Energy under grant Nos. DE-FG02-08ER41533
and DE-FC02-07ER41457 (UNEDF, SciDAC-2), the Research Corporation, the Hungarian Research and Technology Fund, grant No. KTIA AIK 12-1-2012-0020, and the OTKA under grant Nos. T049837 and K101328 is gratefully acknowledged. The authors also would like to thank the anonymous reviewer for valuable comments and suggestions.

\bibliographystyle{apsrev4-1}
\bibliography{paperRefs}

\end{document}